\documentclass[usenatbib]{mn2e}
\usepackage{epsfig}

\title{Evolution of the Luminosity Function and Colours of Galaxies 
in a $\Lam$ Cold Dark Matter Universe}

\author[K. Nagamine et al.]
  {K.~Nagamine,$^1$\thanks{Present address: Harvard College Observatory, 
	60 Garden Street, MS 51, Cambridge, MA 02138, U.S.A.}
	\thanks{Email: nagamine@astro.princeton.edu}  
   M.~Fukugita,$^{2,3}$ R.~Cen,$^4$ J.~P.~Ostriker,$^4$ \\
  $^1$Joseph Henry Laboratories, Physics Department, 
	Princeton University, Princeton, NJ 08544, USA \\
  $^2$Institute for Cosmic Ray Research, University of Tokyo, 
	Kashiwa 2778582, Japan\\
  $^3$Institute for Advanced study, Princeton, NJ 08540, USA\\
  $^4$Princeton University Observatory, Princeton, NJ 08544, USA}

\pagerange{\pageref{firstpage}--\pageref{lastpage}}
\pubyear{2001}

\def\expec#1{\langle#1\rangle}
\newcommand{\Ldenb}{{\cal L}_B}
\newcommand{\Ldenk}{{\cal L}_K}
\newcommand{\Mstarb}{M^{\ast}_B}
\newcommand{\Lstarb}{L^{\ast}_B}
\newcommand{\phistar}{\phi^{\ast}}
\newcommand{\Lam}{\Lambda}
\newcommand{\hinv}{{h^{-1}}}
\newcommand{\mpc}{{\rm\,Mpc}}
\newcommand{\himpc}{\hinv{\rm\,Mpc}}
\newcommand{\hikpc}{\hinv{\rm\,kpc}}
\newcommand{\Msun}{M_{\odot}}

\newcommand{\Lsunb}{L_{\odot,B}}

\newcommand{\himsun}{\hinv{\Msun}}
\newcommand{\etal}{et~al.}

\newcommand{\eg}{{\frenchspacing e.g.}}

\def\Fig#1{Figure~\ref{#1}}

\begin{document}

\maketitle

\label{firstpage}


\begin{abstract}
The luminosity function of galaxies is derived from a cosmological
hydrodynamic simulation of a $\Lam$ cold dark matter (CDM) universe with 
the aid of a stellar population synthesis model.
At $z=0$, the resulting $B$ band luminosity function has a flat faint 
end slope of $\alpha \approx -1.15$ with the characteristic luminosity and 
the normalization in a fair agreement with observations,
while the dark matter halo mass function is steep with a slope of 
$\alpha \approx -2$. The colour distribution of galaxies also agrees well 
with local observations. We also discuss the evolution of the 
luminosity function, and the colour distribution of galaxies from 
$z=0$ to 5. A large evolution of the characteristic mass in the 
stellar mass function due to number evolution is compensated by 
luminosity evolution; the characteristic luminosity increases only 
by 0.8 mag from $z=0$ to 2, and then declines towards higher 
redshift, while the $B$ band luminosity density continues to increase 
from $z=0$ to 5 (but only slowly at $z>3$). 
\end{abstract}



\section{Introduction}
\label{intro_section}

The CDM model provides us with a basis of our understanding of 
cosmic structure formation and galaxy formation \citep{Blumenthal84, 
Davis85}. Much of the recent observational evidence
points to a CDM universe dominated by a cosmological constant 
$\Lam$ \citep{Efstathiou90, Ostriker95, Turner97, Perlmutter98, 
Riess98,     Balbi00, Lange01, Hu01}.  
The physics of galaxy formation, however, is substantially more complicated
than the formation of large-scale structure for which an accurate
treatment of gravity suffices. Many authors use the so-called 
semi-analytic models, in which the dark matter (DM) halo formation 
is supplemented with simple models of dissipative physics for baryons.
Following the pioneering work of \citet{White91}, 
a series of the work \citep{Baugh98, Kauffmann98, SP98, Cole00}
has shown that the model can account for many of the observed 
galaxy properties.

An alternative approach is to use cosmological hydrodynamic 
simulations directly. The advantage is that many physical processes
are automatically taken into account with fewer model assumptions. 
The disadvantage, on the other hand, is that the method is 
computationally expensive, and the resolution is limited by  
computing power. However, the Eulerian hydrodynamic mesh is 
now approaching $(1000)^3$, and we are perhaps beginning to obtain 
meaningful results on the global properties of galaxies. 

Adaptive Mesh Refinement \citep[\eg,][]{Bryan95,  Kravtsov97} and 
Smoothed Particle Hydrodynamics \citep[\eg,][]{Katz96} 
have virtues intermediate between the two listed approaches.
They solve the hydrodynamic equations directly and have a spatial
resolution better than the Eulerian scheme, but they have a mass 
resolution coarser than the Eulerian scheme adopted in this paper. 
Semianalytic models which are based on dark matter halo merger 
trees in N-body simulations also suffer from mass resolution limits.

In a preceding publication \citep[][hereafter Paper I]{Nagamine01},
we discussed galaxy formation history, with a special emphasis 
given to the star formation history and the stellar metallicity distribution. 
In this article, we focus our attention to the luminosity function (LF) 
and its evolution, which allows comparison with one of the most 
fundamental aspects of galaxy observations. We supplement our 
cosmological simulation with a population synthesis model GISSEL99 
\citep[][; Charlot 1999, private communication]{BC93}, which takes 
metallicity variations into account. We assume that stars form from gas 
as soon as the cooling conditions are satisfied.
The \citet{Salpeter55} initial mass function (IMF) with 
a turnover at low-mass end reported by \citet[][hereafter GBF]{Gould96} 
is assumed. The results are presented in the standard Johnson-Morgan 
photometric system. 


\section{Simulation and Parameters}
\label{simulation_section}

The hydrodynamic cosmological simulation we use in this paper 
is the same as that used in Paper I; the comoving box size is 
25$\himpc$, with $768^3$ grid cells and $384^3$ dark matter 
particles each of mass $2.03\times10^7\himsun$. 
The comoving cell size is $32.6\hikpc$, and the mean baryonic 
mass per cell is $3.35\times10^5\himsun$. The cosmological 
parameters are chosen to be $(\Omega_m, \Omega_{\Lam}, 
\Omega_b h^2, h, \sigma_8)=(0.3, 0.7, 0.016, 0.67, 0.9)$.
The basic structure of the code is similar to that of 
\citet{CO92a, CO92b}, but significantly improved over the years
\citep[see][]{CO00}.

We also use a simulation with $384^3$ grid cells (factor 2 and 8 
lower spatial and mass resolution, respectively) to assess the 
effect of the resolution on our result. We call the higher 
resolution run as ``N768'', and the other ``N384''. The input 
parameters are kept to be same for the two runs except the resolution. 

The details of the star formation recipe, the energy feedback,
and the metal production treatments are described in Paper I,
so we do not repeat them here.
The adopted yield $Y=0.02$ \citep{Arnett96} dominantly controls
the resulting amount of stars ($\Omega_{\ast}=0.0052$ including
the 25\% ejected gas by supernovae) and the metal density, which 
are consistent with the upper limit of the empirical estimate 
of \citet{Fukugita98}. We identify galaxies and DM halos using 
the HOP grouping algorithm \citep{Eisenstein98} with the same 
threshold parameters as in Paper I. We refer to the grouped stellar 
particles as `galaxies' hereafter. 
We confirmed that all galaxies identified by HOP are dynamically 
stable enough to obtain meaningful results for the evolution 
of individual galaxies (see Paper I).


\section{Mass and Luminosity Function at $z=0$}
\label{zzero_section}

\begin{figure}
\epsfig{file=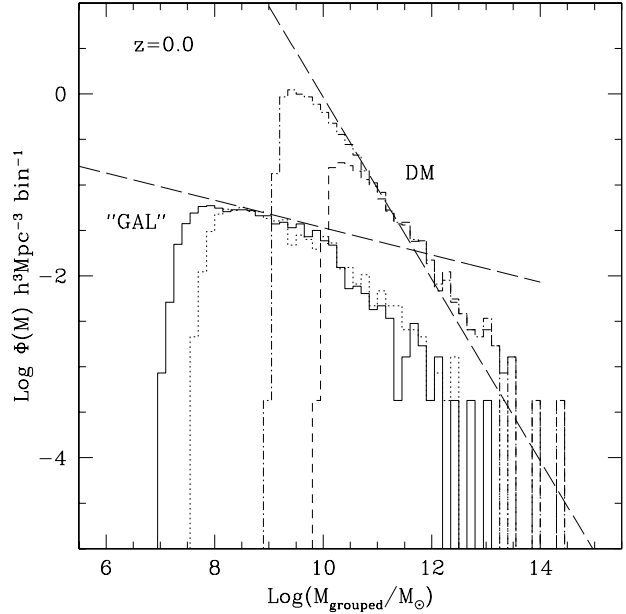,height=3.2in,width=3.2in, angle=0} 
\caption{Mass function of dark matter halos (dash-dotted N768 \& 
short-dashed N384 histogram) and 
galaxy stellar mass (solid N768 \& dotted N384 histogram).
The two long-dashed lines show 
$\phi(M)\sim M^{-2}$ and $\phi_g(M)\sim M^{-1.15}$.}
\label{fig1.eps}
\end{figure}

In \Fig{fig1.eps}, we show the DM halo mass function $\Phi(M)$ 
in dash-dotted (N768) and short-dashed (N384) histograms.
The galaxy stellar mass function $\Phi_g(M)$ is also shown
in solid (N768) and dotted (N384) histograms at $z=0$,
where $\Phi(M)d\log(M)\equiv \phi(M)dM$ (likewise for the LF below). 
The two long-dashed lines show 
$\phi(M)\sim M^{-2}$ and $\phi_g(M)\sim M^{-1.15}$.

The DM halo distribution follows $\phi(M)\sim M^{-2}$ well, which is 
the generic mass distribution of the hierarchical clustering model. 
Some rounding at the low mass end is seen below $10^{10}\Msun$ 
(500 DM particles), and a slight overproduction of massive halos 
is seen at above $10^{13}\Msun$ due to overmerging problem. 

The galaxy stellar mass function behaves quite differently: 
flattening below $10^{10}\Msun$ is apparent, which
can be ascribed to supernova feedback \citep{Dekel86} and 
photoheating of gas \citep{Rees86, Efstathiou92},
both more efficient in less massive galaxies. 
The results of N768 and N384 agree well in the overlapping mass range
for both the dark matter and the galaxy stellar mass functions, 
verifying that the results that concern us in this paper are not 
strongly affected by the resolution effects. 
Therefore, only N768 run is used in the rest of this paper.

At above $3\times 10^{10}\Msun$, the galaxy stellar mass 
function departs from the power law and seems to follow DM halo 
mass function. This departure from the faint-end power law can
partly be ascribed to the inefficient cooling in massive galaxies 
\citep{Rees77, Silk77}.
The cutoff at high-mass end, however, is not quite exponential; 
there are several objects with a very large stellar mass of
$>5\times 10^{11}\hinv\Msun$ (indicated by the dotted histogram),
due to an overmerging problem. These objects have the mass of groups 
and clusters, and the galaxies have overmerged in very high-density 
regions due to the limited spatial resolution of the simulation.
There is no good way to decompose the overmerged objects into 
individual galaxies, and it is difficult to estimate the effect of 
the overmerging to the shape of the high-mass end of the stellar 
mass function. 

\begin{figure}
\epsfig{file=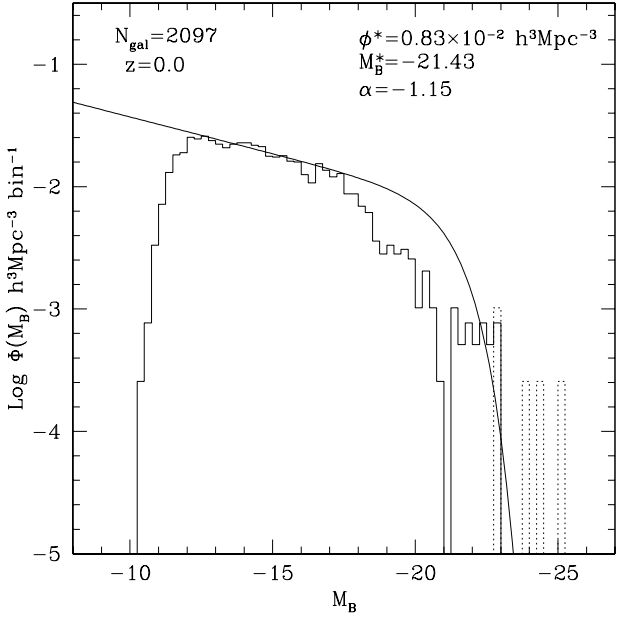,height=3.2in,width=3.2in, angle=0} 
\caption{Luminosity function of galaxies at $z=0$.
The galaxies with stellar mass of $>5\times10^{11}\hinv\Msun$
are shown as dotted histogram, and the solid curve is 
the Schechter function with the indicated parameters in the 
figure. See text for the method of the fitting.}
\label{fig2.eps}
\end{figure}

The LF in the rest-frame $B$ band (before dust extinction) is 
presented in \Fig{fig2.eps} as a function of absolute magnitude 
(the overmerged objects with stellar mass $>5\times10^{11}\hinv\Msun$ 
are indicated by the dotted histograms).
We attempt to fit the Schechter function to the computed LF in the 
following manner: the faint end slope $\alpha$ is fixed to $-1.15$ 
throughout this paper, we then choose $\phistar$ to fit the faint 
end plateau of the LF, and adjust $\Mstarb$ so that the chosen 
Schechter function reproduces the computed luminosity density when 
it is integrated over.
Although the overmerging problem hampers accurate determination of 
the Schechter parameters of the computed LF, they can still be 
used to characterize our result with a minimum number of adjustable
parameters. 
In the figure, the Schechter function with $\alpha=-1.15$, 
$\Mstarb=-21.43$, and $\phistar=0.83\times10^{-2}h^3\mpc^{-3}$ is 
shown in the solid curve. The simulated LF fits the slope of 
$\alpha=-1.15$ well at fainter magnitudes below $M_B=-17$.
If the overluminous objects are not included in the fit, 
then we obtain $\Mstarb=-20.68$, which is 0.75 magnitude 
dimmer than the case of full luminosity. 
This may suggest that our $\Mstarb$ at $z<1$ is biased towards 
the brighter side because of the overmerged objects.

The computed $B$ band luminosity density $\Ldenb$ of galaxies 
at $z=0$ is $2.4\times 10^8h\Lsunb\mpc^{-3}$ including the 
overmerged objects. 
A source of large uncertainty in the luminosity density is the 
assumed IMF. The use of Scalo IMF instead of the Salpeter IMF 
would increase the computed $\Ldenb$ by $\sim 30$\% \citep{White91}.
Another source of uncertainty is the dust extinction. 
The empirical average value of extinction in the $B$ band 
is $A_B=0.33$ mag \citep{Vaucouleurs91}.
The corresponding decrease in $\Ldenb$ by dust extinction is by 30\%.

Optical surveys of galaxies have made significant progress
over the last 10 years. In particular, the 2dF Survey \citep{2dF} 
and the Sloan Digital Sky Survey \citep{Blanton01, Yasuda01}
have secured the luminosity functions and the local luminosity density of
the universe. Both surveys agree on $\alpha=-1.20\pm0.10$ and 
$M^*_B=-19.8\pm0.2+5\log h$, and the luminosity density 
$\Ldenb=2.4\pm0.4\times10^8h\Lsunb$. 
Our $\Mstarb$ at $z=0$ is brighter than empirical values by
0.76 magnitude, and $\phistar$ is lower by a factor of $\sim 2$.
Overmerged objects contribute about 50\% of the total luminosity 
in the box, but the overmerging does not affect the total luminosity
density, as the fraction of baryons that condense into stars is 
basically determined by the yield parameter in the simulation
(see Paper I).

The uncertainty in the fitted value of $\Mstarb$ associated with 
the overmerging, IMF, and dust extinction is of the same order 
as the above discrepancy between the computed and the observed LF. 
In view of the fact that we have not fine-tuned the simulation 
parameters, we consider the agreement of the computed 
LF and $\Ldenb$ with observations is fair.


\section{Evolution of Luminosity Function}
\begin{figure}
\epsfig{file=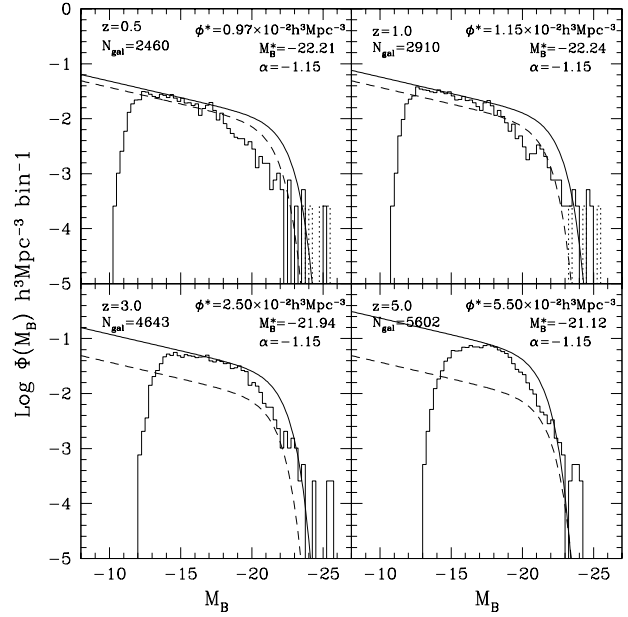,height=3.2in,width=3.2in, angle=0} 
\caption{Luminosity function of galaxies at $z=0.5, 1, 3,$ and 5.
The `galaxies' with stellar mass $>5\times10^{11}\hinv\Msun$ 
are shown as dotted histograms. The solid curves are the Schechter
functions with the indicated parameters in each panel, and the
dashed curve is the Schechter fit to the simulated LF 
at $z=0$ shown in \Fig{fig2.eps}.}
\label{fig3.eps}
\end{figure}

\Fig{fig3.eps} shows the rest-frame $B$ band LF
at $z=0.5, 1, 3,$ and 5. The solid curves are the Schechter 
functions with the indicated parameters (see also Table 1), 
chosen in the same manner as in \Fig{fig2.eps}. 
The dashed curve is the Schechter function at $z=0$ for comparison. 

We observe some evolution between $z=0$ and 1. 
The characteristic magnitude $\Mstarb$ brightens by 0.8 mag,
and the normalization $\phistar$ increases by a factor of 1.4.
As a result, $\Ldenb$ at $z=1$ is 2.9 times higher than at $z=0$ (Table 1). 
The faint end slope evolves very little. 

We find that the evolution of the LF is different from that of 
the galaxy stellar mass function, which is also fitted well by the 
Schechter function except at the high-mass end.
The characteristic mass $M^*_g$ of the galaxy stellar mass function 
decreases by a factor of 2.4 from $z=0$ to 1, 
but the number of galaxies increases by a factor of 1.4, 
leading to a moderate decrease in the stellar mass density $\rho_*$ 
by a factor of 1.5 (Table 1).
These changes are compensated by the luminosity evolution:
the average $B$ band mass-to-light ratio of galaxies decreases by a factor
of 2.4, leading to more luminous $\Lstarb$ at higher redshift.
The increase of luminosity to higher redshift is due to increased 
star formation of blue galaxies (Paper I), but it is also partly due 
to the passive evolution of red galaxies, as evidenced from the shift 
of the red edge in the colour distribution, as we see below.

\begin{table*}
 \centering
 \begin{minipage}{140mm}
  \caption{GALAXY STATISTICS PARAMETERS \label{table1}}
  \begin{tabular}{@{}cccccccc@{}}
Quantities & $z=0$    & $z=0.3$  & $z=0.5$  & $z=1$    & $z=2$    & $z=3$    & $z=5$ \\
$\Mstarb$  & $-21.43$ & $-21.93$ & $-22.21$ & $-22.24$ & $-22.27$ & $-21.94$ & $-21.12$ \\
$\phistar$ & 0.83     & 0.93     & 0.97     & 1.15     & 1.60     & 2.50     & 5.50 \\
$\Ldenb$   & 1.62     & 2.86     & 3.86     & 4.69     & 6.74     & 7.77     & 8.02 \\ 
$\expec{M/L_B}$ & 4.3 & 3.2      & 2.7      & 1.8      & 0.87     & 0.44     & 0.16 \\
$\expec{M/L_K}$ & 1.3 & 1.0      & 0.91     & 0.69     & 0.42     & 0.30     & 0.21 \\
$\rho_*/\Ldenb$ & 4.0 & 2.1      & 1.4      & 0.94     & 0.44     & 0.25     & 0.13 \\
$\rho_*/\Ldenk$ & 0.77 & 0.53     & 0.49     & 0.39     & 0.27     & 0.21     & 0.15 \\
$\expec{B-V}$ & 0.70  & 0.68     & 0.66     & 0.61     & 0.50     & 0.36     & 0.12 \\ 
$\expec{U-B}$ & 0.09  & 0.07     & 0.06     & 0.05     & 0.03     & -0.03    & -0.28 \\ 
$\expec{V-K}$ & 2.29  & 2.23     & 2.18     & 2.09     & 1.92     & 1.68     & 1.20 \\ 

$M_{g}^*$  & 2.34     & 1.97     & 1.45     & 0.97     & 0.41     & 0.16     & 0.05 \\
$\phistar_{g}$ & 0.83 & 0.93     & 1.10     & 1.35     & 2.20     & 3.60     & 6.00 \\
$\rho_*$   & 6.50     & 6.13     & 5.34     & 4.39     & 3.00     & 1.98     & 1.03 \\ 
\end{tabular}

\medskip
Parameters of the Schechter functions shown in \Fig{fig2.eps} and \ref{fig3.eps} 
are presented in the first two rows.
$\alpha$ is fixed to $-1.15$ in all cases. See text for the method 
of choosing the values of $\Mstarb$ [mag] and $\phistar [10^{-2}h^3\mpc^{-3}]$.
$\Ldenb$ is in units of [$10^8\Lsunb\mpc^{-3}$] with $h=0.67$.
The average mass-to-light ratio of all galaxies in solar units, and the 
average of the entire box (mass density divided by luminosity density) are also given. 
The Schechter parameters of the galaxy stellar mass function ($M_{g}^* [10^{11}\Msun]$ 
and $\phistar_{g} [10^{-2}h^3\mpc^{-3}]$) and the comoving stellar mass density 
$\rho_{\ast}[10^8\Msun\mpc^{-3}]$ are shown in the bottom three rows.
\end{minipage}
\end{table*}

Empirical knowledge of the evolution of the LF is
still controversial, despite much effort made by the Canada-France
Redshift Survey \citep[CFRS;][]{CFRS}, 
the Autofib Redshift Survey \citep{Autofib},
the CNOC2 Survey \citep{CNOC2}, 
the CADIS Survey \citep{CADIS},
and the DEEP project \citep{DEEP}.
The CFRS and the CADIS claim that the blue population brightens by 
about 1 magnitude from $z=0$ to 0.8, while the characteristic
luminosity changes little for the red population. 
On the other hand, the CNOC2 and the DEEP claim brightening of 
characteristic luminosity for red (or early) populations.
The Autofib indicates little evolution of $\Lstarb$
for total luminosity function, while it claims steepening of the 
faint end slope. 

Overall, we summarize that the change in $\Lstarb$ 
from $z=0$ to 1 deduced from the observations for the total LF 
is modest, no more than $\sim 1$ mag 
in the $\Lam$ cosmology, whereas the appreciable increase in 
luminosity density is common to all surveys.
The values of luminosity density are discrepant among authors, 
and their low-redshift values are also not consistent 
with what we referred to for $z=0$ above. 
By compiling all results, we conclude that the increase
of $\Ldenb$ from $z=0$ to 0.5 is by a factor of $1.4-2$,
and by a factor of $2-3$ from $z=0$ to 1. 
The result from our simulation is consistent 
with these data at the upper end for both $\Lstarb$ and $\Ldenb$.
We have not seen steepening of the total LF towards higher redshift 
in our simulation.

The trend of evolution is somewhat different when one goes to higher 
redshift. The characteristic mass of the galaxy stellar mass function
decreases more rapidly from $z=1$ to 3 (by a factor of 6).
This change is faster than that of the mass-to-light ratio,
resulting in a moderate decrease of the characteristic luminosity. 
The normalization of the LF increases rapidly towards higher redshift, 
but the increase in $\Ldenb$ is smaller above $z=2$,
because galaxies become less massive and dimmer at the same time.   
 
We find that the $U$-band luminosity density continues to increase
from $z=0$ to 5, whereas the $K$-band luminosity density gradually decreases
towards high-redshift, at least from $z=2$ (some increase is
observed from $z=0$ to 1). 


\section{Galaxy Colours and their Evolution}
\label{colour_section}
\begin{figure}
\epsfig{file=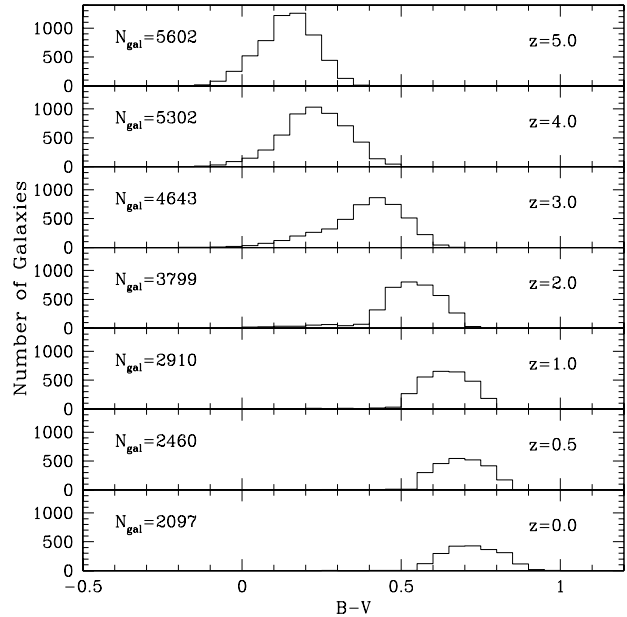,height=3.2in,width=3.2in, angle=0} 
\caption{Rest-frame $B-V$ colour distribution of galaxies at various 
redshifts. See text for discussions.}
\label{fig4.eps}
\end{figure}

\Fig{fig4.eps} presents the $B-V$ colour distribution of galaxies 
at various redshifts. 
At $z=0$, the $B-V$ colour ranges from 0.55 to 0.95 with a tiny  
fraction of galaxies (0.6\% in number, but 12\% in luminosity) 
bluer than 0.5 that represents very active star-forming galaxies.
The median $B-V$ is 0.70. 
These results are in global agreement with observations 
for local galaxies, which ranges from $B-V=0.45$ (the bluest Im) 
to 1 (giant E); $B-V=0.7$ is the median colour of Sb galaxies 
\citep*{Buta94, Fukugita95}. 
In finer details, the calculated distribution is slightly narrower 
than the observations in both blue and red edges.

As one goes to higher redshift, the red edge is shifted, which is 
ascribed to the younger age of galaxies, since the red edge is mainly 
determined by the maximal available time for passive evolution of galaxies
up to each epoch. 
The shape of the distribution on the red side does not change
very much as a function of redshift. 
Another conspicuous fact as one goes to higher redshift is the 
increasingly longer tail towards the blue side, indicating 
star formation in larger fractions of galaxies.
At high-redshift, this star-forming population rapidly increases, 
and at $z=4$ the bulk of the population is star-forming.
In addition to the increasing star formation, 
the stellar population tends to be bluer due to 
lower metallicity at high-redshift (see Paper I).

The $U-B$ colour at $z=0$ ranges from $-0.05$ to 0.4 in good 
agreement with local observations, except that it is a little 
narrower than the observations \citep{Buta94, Fukugita95}.
The evolution is qualitatively similar to $B-V$.
Assuming a typical extinction parameter of $R_V=3.1$ and the extinction 
law of \citet{Cardelli89}, one finds that the $U-B$ colour 
becomes redder by 0.06 mag.

The $V-K$ colour at $z=0$ ranges from 1.9 to 2.9.
This is perhaps bluer by 0.1 mag at 
the red edge, and 0.3 mag bluer at the blue edge 
than the real world \citep{Fioc99}. 
This bluer colour is ascribed to the GISSEL99 model, 
with which $V-K$ colour 1 Gyr after the burst is 
only 2.0, and that at 14 Gyrs is 3.0 for a galaxy 
with 40\% solar metallicity (see Paper I for 
the metallicity of galaxies in our simulation).

\section{Discussion}
\label{discussion_section}
Our $\Lam$CDM simulation reproduces the global statistics concerning 
the luminosity and colours of galaxies fairly well at $z=0$ considering
the uncertainties in the model and the simulation.
Even the current level of agreement is a significant improvement
over the previous simulations, and we are satisfied with the
qualitative agreement for the moment, given that we have not done 
the fine-tuning of the input parameters.

The number evolution and the luminosity evolution of galaxies takes 
place at the same time in the simulation. The galaxies in the simulation 
continues to become less massive and more numerous from $z=0$ to 5, 
as expected in the hierarchical structure formation scenario, but 
luminosity per stellar mass significantly increases to high redshifts.
Because of the compensation of the two effects, the change in the LF is 
modest: the characteristic luminosity increases from $z=0$ to 2
only by 0.8 mag, and then decreases beyond $z>2$. The evolution of the LF
is a modest representation of the evolution of individual galaxies.
The $B$ band luminosity density continues to increase from $z=0$ to 5, 
but only slowly at $z>3$. We do not see the apparent steepening of 
the faint end slope of the LF.
Uncertainties in the observational LFs at non-zero redshifts do not allow
a finer comparison of the model prediction with the data.

In detailed level the agreement between the predictions and the observations 
is not perfect. However, given the uncertainties in the theoretical 
modelling and observations, the level of agreement presented in this 
paper is encouraging.
The qualitative evolutionary trends presented in this paper 
(and those in Paper I) can be taken as predictions of hydrodynamic 
simulations based on the $\Lam$CDM model plus the prescriptions 
for star formation and feedback effects. These qualitative predictions 
can be used to test the $\Lam$CDM scenario against further observational 
studies.


\section*{Acknowledgments}

We thank Michael Strauss for useful comments on the draft.
K.N. is supported in part by the Physics Department.
M.F. is supported in part by the Raymond and Beverly Sackler Fellowship 
in Princeton and Grant-in-Aid of the Ministry of Education of Japan.
R.C. and J.P.O. are partially supported by grants AST~98-03137 and 
ASC~97-40300.


\bsp

\label{lastpage}

\end{document}